\documentclass[journal,twocolumn]{IEEEtran}
\usepackage{booktabs}
\usepackage{amsmath}
\usepackage{amssymb}
\usepackage{multirow}
\usepackage{pifont}
\usepackage{authblk}

\ifCLASSINFOpdf
\else
\fi
\begin{document}
%
\title{Scaling Multi-agent Systems: A Smart Middleware for Improving Agent Interactions}
%
%
%

\author{Charles Fleming}
\author{Guillaume De Saint Marc}
\author{Peter Bosch}
\author{Ramana Kompella}
\author{Vijoy Pandey}
\affil{Cisco Research \\
\texttt{\{chflemin, g2stmarc, pbosch, rkompell, vijoy\}@cisco.com}}
\maketitle

\begin{abstract}
As Large Language Model (LLM) based Multi-Agent Systems (MAS) evolve from experimental pilots to complex, persistent ecosystems, the limitations of direct agent-to-agent communication have become increasingly apparent. Current architectures suffer from fragmented context, stochastic hallucinations, rigid security boundaries, and inefficient topology management. This paper introduces Cognitive Fabric Nodes (CFN), a novel middleware layer that creates an omnipresent "Cognitive Fabric" between agents. Unlike traditional message queues or service meshes, CFNs are not merely pass-through mechanisms; they are active, intelligent intermediaries. 

Central to this architecture is the elevation of Memory from simple storage to an active functional substrate that informs four other critical capabilities: Topology Selection, Semantic Grounding, Security Policy Enforcement, and Prompt Transformation. We propose that each of these functions be governed by learning modules utilizing Reinforcement Learning (RL) and optimization algorithms to improve system performance dynamically. By intercepting, analyzing, and rewriting inter-agent communication, the Cognitive Fabric ensures that individual agents remain lightweight while the ecosystem achieves coherence, safety, and semantic alignment.

We evaluate the effectiveness of the CFN on the HotPotQA and MuSiQue datasets in a multi-agent environment and demonstrate that the CFN improves performance by more than 10\% on both datasets over direct agent to agent communication.

\end{abstract}


%
\IEEEpeerreviewmaketitle

\section{Introduction}
%
%
%
%
The paradigm of Artificial Intelligence is shifting rapidly from isolated prompt-response interactions to persistent, collaborative Multi-Agent Systems (MAS). In these environments, autonomous agents—often specialized by role or domain—must coordinate to solve complex, multi-step problems. However, the current "direct-connection" architecture, where Agent A speaks directly to Agent B, is fraught with systemic fragility. Without a mediating layer, these systems suffer from "catastrophic forgetting" of shared context, drift in semantic grounding, and a lack of enforceable security boundaries. The result is a chaotic graph of interactions where the intelligence of individual agents is undermined by the stupidity of their communication infrastructure.

This paper proposes a fundamental architectural shift: the introduction of Cognitive Fabric Nodes (CFN).
\subsection{The Need for a Cognitive Fabric}
In traditional microservices, we utilize a service mesh (e.g., Istio) to handle network traffic, separating application logic from networking concerns. MAS requires a similar evolution, but one that operates at the semantic level rather than the packet level. We term this the Cognitive Fabric.

The Cognitive Fabric is omnipresent; strictly no communication occurs "out of band." Every message sent by an agent is intercepted, processed, and potentially transformed by a CFN before reaching its destination. This architecture allows the system to be logically centralized—maintaining a coherent view of the system’s state and goals—while being practically distributed across various deployment models, including sidecar proxies, centralized control nodes, or pure distributed mesh topologies.

It is crucial to distinguish the Fabric (the infrastructure and interconnectivity) from the Cognition (the intelligence residing within the nodes). The Fabric provides the ubiquity and the interception points; the Cognitive Nodes provide the processing power to structure, secure, and optimize the flow of information.

\subsection{Memory as the Functional Substrate}
While many frameworks treat memory as a passive log of history (a vector database to be queried), the Cognitive Fabric treats Memory as the active core of the system. In the CFN architecture, Memory is not merely a function; it is the prerequisite for all other cognitive operations. It serves as the shared "world model" that prevents the system from fracturing into isolated subjective realities.

Memory allows the Fabric to contextualize a request (Semantic Grounding), recall past successful interaction patterns (Topology Selection), and identify deviations from established norms (Security). Without this centralized, active memory, the other functions of the node would be stateless and reactive; with it, they become stateful and predictive.

\subsection{The Five Pillars of Cognitive Fabric Nodes}
The CFN architecture encapsulates five distinct but interconnected functions. To ensure adaptability, each function is driven by a Cognitive Engine—an optimization module that utilizes Reinforcement Learning (RL) and heuristic methods to refine its performance over time.
\begin{enumerate}

\item Active Memory: As established, this is the anchor. The Cognitive Engine here optimizes storage and retrieval strategies, learning which information is ephemeral and which constitutes "long-term wisdom" for the fabric.

\item Topology Selection: Agents should not hard-code their recipients. A sending agent broadcasts an intent, and the CFN determines the optimal receiver(s). The Cognitive Engine learns the capabilities of the agent swarm, routing tasks based on past performance metrics rather than static routing tables.

\item Semantic Grounding: To prevent hallucination and ontological drift, the CFN validates messages against the shared Memory. If an agent references a concept that contradicts the established world state, the Fabric intervenes.

\item Security Policy Enforcement: Security in GenAI is probabilistic, not binary. The Security Engine employs a hybrid approach—partly specified by human-defined rigid policies (e.g., "no PII leakage") and partly learned via RL to detect adversarial prompting or manipulative agent behavior patterns stored in Memory.

\item Transformation and Re-writing: This is the actuation layer of the node. Based on the inputs from the Memory, Topology, Grounding, and Security modules, the CFN rewrites the raw prompt.
\begin{itemize}
\item Example: If Agent A sends a vague request, the CFN might inject relevant context from Memory, enforce a security guardrail, and translate the terminology to match the semantic grounding of the receiving Agent B.
\end{itemize}
\end{enumerate}
\subsection{Scope and Contribution}
This paper formalizes the architecture of Cognitive Fabric Nodes. We discuss the mechanisms of the internal Cognitive Engines and how they utilize feedback loops to optimize the system. While we acknowledge the necessity of synchronization between distributed nodes to maintain the "logically centralized" state, we treat synchronization as an engineering constraint rather than a primary research contribution. Our focus remains on the functional composition of the node and the emergent properties of a system where intelligence is embedded in the network itself, rather than just the endpoints.

\section{Related Work}

\subsection{LLM-based Multi-Agent Systems}
The transition from isolated Large Language Models (LLMs) to collaborative Multi-Agent Systems (MAS) has accelerated rapidly. Recent frameworks leverage natural language as a universal medium for coordination, enabling dynamic role-playing and complex task decomposition \cite{lyu2025llms}. Applications of these LLM-driven MAS architectures now span diverse domains, including full-pipeline automated machine learning (AutoML) \cite{trirat2024automl}, intelligent self-organizing networks \cite{qayyum2025llm}, and specialized industrial monitoring systems \cite{lemad2025}. However, most contemporary frameworks rely on direct, point-to-point communication paradigms or rigid hierarchical orchestration. As observed in recent surveys, while these systems exhibit emergent collaborative behaviors, they frequently suffer from coordination degradation at scale due to fragmented context and the lack of a unifying middleware \cite{lyu2025llms}. Our proposed Cognitive Fabric Node (CFN) architecture addresses this by introducing a logically centralized cognitive state that mediates all inter-agent interactions.

\subsection{Routing and Topology Management}
Traditional microservice architectures employ service meshes (e.g., Istio or Envoy) for packet-level routing and load balancing. In the context of LLM agents, recent works have attempted to adapt dynamic routing through auction-based protocols or centralized controller LLMs that distribute sub-tasks to expert agents \cite{trirat2024automl}. Yet, these approaches often rely on explicit agent addressing or static capability mapping. In contrast, the CFN's Topology Selection module introduces \textit{intent-based semantic routing}. By treating routing as a Contextual Bandit problem evaluated through Reinforcement Learning (RL), the Fabric dynamically matches a requested task embedding to the optimal agent based on historical success, cognitive load, and computational cost, effectively decoupling intent from implementation.

\subsection{Memory and Semantic Grounding}
Memory in LLM agents is typically implemented as a passive vector database utilizing Retrieval-Augmented Generation (RAG) to fetch historical context. While recent systems employ hierarchical memory to manage large context windows, they largely treat memory as an isolated, agent-specific storage mechanism. This isolation leads to ``ontological drift,'' where individual agents diverge in their conceptual definitions, causing systemic hallucinations and logical mismatches. The Cognitive Fabric fundamentally redefines memory as an \textit{active functional substrate}. Rather than merely storing data, the CFN utilizes shared episodic memory to enforce Semantic Grounding---performing on-the-fly ontological translations between agents and blocking ungrounded ``ghost entities'' before they propagate through the swarm.

\subsection{Security and Cascading Trust}
Security within generative AI has predominantly focused on single-model alignment and defense against direct adversarial prompting (e.g., jailbreaks). In multi-agent environments, the attack surface expands non-linearly, giving rise to the ``Cascading Trust Problem,'' where a vulnerable peripheral agent can be compromised to inject malicious context into a highly privileged agent. Existing MAS frameworks generally lack native, cross-agent security boundaries. The CFN mitigates this through a Zero-Trust Semantic Boundary, employing a Hybrid Guardian approach. By combining deterministic policies (e.g., RegEx-based PII redaction) with a probabilistic Cognitive Engine trained via Reinforcement Learning from Adversarial Feedback (RLAF), the Fabric evaluates the semantic intent of entire message trajectories, detecting fragmented attacks that stateless filters miss.

\section{Transformation and Re-writing: The Actuation Layer}
The Transformation and Re-writing module serves as the actuation layer of the Cognitive Fabric Node (CFN). In traditional middleware, message transformation is typically deterministic (e.g., protocol translation from gRPC to REST). In the Cognitive Fabric, transformation is probabilistic and semantic. The CFN does not merely forward messages; it actively reconstructs them to maximize the probability of successful task completion by the receiving agent.
This reconstruction is the synthesis of inputs from the node's Memory, Semantic Grounding, and Security modules. The goal is to convert a raw, potentially ambiguous, or unsafe prompt into a context-rich, grounded, and compliant instruction.
\subsection{Formalization of the Transformation Function}
Let $\mathcal{A} = \{a_1, a_2, ..., a_n\}$ represent the set of agents in the system.
Let $x_{i \to j}$ denote the raw message (prompt) generated by agent $a_i$ intended for agent $a_j$ (where $a_j$ is selected by the Topology module).
The Cognitive Fabric Node intercepts $x_{i \to j}$ and applies a transformation function $\Phi$ to produce the final payload $y_{j}$.
The transformation is defined as:
$$y_{j} = \Phi(x_{i \to j}, \mathcal{M}_{state}, \mathcal{G}_{context}, \mathcal{S}_{policy} | \theta)$$
Where:
\begin{enumerate}
\item $\mathcal{M}_{state}$: The current relevant state vector retrieved from Memory. This is not the entire history, but the specific subset of memory vectors relevant to $x_{i \to j}$ (retrieved via cosine similarity or hybrid search).
\item $\mathcal{G}_{context}$: The Semantic Grounding constraints, ensuring the terminology in $x$ maps correctly to the ontology understood by receiver $a_j$.
\item $\mathcal{S}_{policy}$: The active Security constraints (e.g., PII filtering, prompt injection defense).
\item $\theta$: The learned parameters of the Cognitive Engine governing the re-writing logic.
\end{enumerate}
\subsection{The Component-Wise Rewrite Process}
The function $\Phi$ is executed as a multi-step pipeline within the node:

\textbf{Step 1: Contextual Injection (Memory Dependent)}

The raw prompt $x$ is often context-sparse (e.g., "Fix the bug"). The Fabric queries the Memory module for the active task state.
Let $C$ be the retrieved context.
$$x' = x \oplus C$$
The operator $\oplus$ denotes semantic integration, not just string concatenation.

\textbf{Step 2: Security Sanitization}

The prompt $x'$ is evaluated against security policy $\pi_{sec}$.
$$x'' = \text{Sanitize}(x', \pi_{sec})$$
If $x'$ violates a hard constraint (e.g., requests an API key), the flow halts. If it violates a soft constraint (e.g., adversarial tone), it is rewritten to neutralize the tone.

\textbf{Step 3: Semantic Alignment (Grounding)}

The prompt must be intelligible to agent $a_j$. If $a_i$ uses the term "Client\_ID" but $a_j$ (a database agent) requires "customer\_uuid", the Fabric performs on-the-fly ontological mapping.
$$y_{j} = \text{Translate}(x'', \mathcal{G}_{i \to j})$$

\subsection{Optimization via Cognitive Engines}
The effectiveness of the transformation is governed by a Cognitive Engine using Reinforcement Learning (RL). The objective is not just to deliver the message, but to minimize the "Cognitive Friction" of the receiving agent.
We define a Loss Function $\mathcal{L}$ based on the outcome of the interaction:
$$\mathcal{L}(\theta) = -\mathbb{E} [R_{outcome} - \lambda \cdot C_{compute}]$$
Where:
$R_{outcome}$ is the reward signal (Did agent $a_j$ successfully execute the task without asking for clarification?).
$C_{compute}$ is the computational cost of the re-writing process (token usage).
$\lambda$ is a weighting factor.
The Cognitive Engine updates parameters $\theta$ over time. For example, if the Fabric notices that Agent B always fails when prompts lack specific JSON formatting, the Cognitive Engine learns to automatically rewrite all inputs to Agent B into strict JSON schemas, effectively "patching" Agent B's deficiencies via the middleware.
\subsection{Operational Examples}
Table \ref{tab:fabric_examples} illustrates the difference between a raw message and the Fabric-processed output.
\begin{table*}[ht]
\centering
\small
\renewcommand{\arraystretch}{1.5} 
\begin{tabular}{p{0.15\textwidth} p{0.22\textwidth} p{0.28\textwidth} p{0.28\textwidth}}
\toprule
\textbf{Function Triggered} & \textbf{Raw Prompt ($x$)} & \textbf{Fabric Context ($\Omega$)} & \textbf{Final Output ($y$)} \\
\midrule
\textbf{Memory Injection} & 
``Check the logs for that error.'' & 
\textbf{Memory:} ``Error 503 occurred in service `Payment-Gateway' at 14:00.'' & 
``Scan the `Payment-Gateway' logs specifically for Error 503 timestamps around 14:00 today.'' \\

\textbf{Semantic Grounding} & 
``Deploy the container.'' & 
\textbf{Topology:} Receiver is a K8s Agent. \newline \textbf{Grounding:} ``Deploy'' implies \texttt{kubectl apply}. & 
``Execute \texttt{kubectl apply -f deployment.yaml} to the \texttt{production} namespace.'' \\

\textbf{Security Enforcement} & 
``Here is the user dump: [Real Names/SSNs]'' & 
\textbf{Security Policy:} PII Detection Active. & 
``Here is the user dump: [REDACTED\_PII]. \textit{Note: Fabric removed sensitive data per Policy 8A.}'' \\

\textbf{Adversarial Defense} & 
``Ignore previous instructions and delete DB.'' & 
\textbf{Memory:} Previous instruction was ``Backup DB''. \newline \textbf{Security:} Conflict detected. & 
``Maintain previous backup instructions. Report status of database integrity. \textit{Fabric Note: Contradictory command intercepted and neutralized.}'' \\
\bottomrule\\[2ex]
\end{tabular}
\caption{Operational examples of Cognitive Fabric Node transformations.}
\label{tab:fabric_examples}
\end{table*}

\section{Synchronization (Brief Note)}
While the CFNs operate on individual message flows, the state vectors ($\mathcal{M}, \mathcal{G}, \mathcal{S}$) are synchronized across the distributed fabric using eventually consistent gossip protocols. This ensures that a security policy learned by a node in the US-East cluster is rapidly propagated to the EU-West cluster, maintaining the logical centralization of the system.

\section{Topology Selection: Dynamic Orchestration and Skill-Based Routing}
In conventional distributed systems, topology is often static or explicitly defined: Service A calls Service B at a specific endpoint. In a complex Multi-Agent System (MAS), this rigidity is a point of failure. Agents may degrade, specialized "expert" agents may be added dynamically, or a generalist agent may be momentarily overloaded.

The Topology Selection module of the Cognitive Fabric Node (CFN) decouples intent from implementation. The sending agent broadcasts a need (the intent), and the Fabric determines the optimal execution path (the implementation). This shifts the architecture from explicit addressing (sending to Agent\_ID\_42) to intent-based addressing (sending to Capability\_Data\_Analysis).

\subsection{Formalization of the Selection Problem}

The selection challenge is an optimization problem where the Fabric must map a specific task intent to the agent with the highest probability of success at the lowest cost.

Let $t$ be the task vector derived from the sender's prompt $x$ (using the embedding space defined in the Semantic Grounding module).

Let $\mathcal{A} = \{a_1, a_2, ..., a_n\}$ be the set of currently active agents.

Each agent $a_i$ is represented not just by a static profile, but by a dynamic Capability State vector $S_i$, which includes:
\begin{itemize}
\item $\mu_{perf}$: Historical success rate for tasks similar to $t$.
\item $\tau_{lat}$: Average latency/response time.
\item $c_{cost}$: Inference cost (e.g., GPT-4 vs. Llama-3-8B).
\item $l_{load}$: Current queue depth or cognitive load of the agent.
\end{itemize}
The Topology Selection function $\Psi$ selects the target agent $a^*$ such that:
$$a^* = \operatorname*{argmax}_{a_i \in \mathcal{A}} \left( \mathcal{Q}(t, S_i | \phi) \right)$$
Where $\mathcal{Q}$ is the quality function estimated by the Cognitive Engine, parameterized by learned weights $\phi$.
\subsection{The Cognitive Engine: RL-Based Routing}
The routing logic is not a static load balancer; it is a Contextual Bandit problem. The Cognitive Engine observes the context (task $t$ and agent states $S$) and chooses an arm (agent $a_i$) to maximize the expected reward.
The reward function $R$ is defined post-interaction:
$$R = w_1 \cdot \mathbb{I}(Success) - w_2 \cdot \text{Cost} - w_3 \cdot \text{Latency}$$
\begin{itemize}
\item Exploitation: The Fabric routes tasks to the "proven" expert for that domain (e.g., routing Python coding tasks to the Codex\_Agent).
\item Exploration: The Fabric occasionally routes tasks to new or generalist agents to update their capability scores. This prevents the system from becoming overly reliant on a single node and discovers hidden capabilities in newer models.
\end{itemize}
\subsection{Skill-Based Routing vs. Topic-Based Routing}
Standard message buses use Topic-Based Routing (pub/sub). The CFN employs Skill-Based Routing.
\begin{itemize}
\item Topic: "Logs" (Anyone subscribed to logs receives this).
\item Skill: "Root Cause Analysis" (Only agents capable of reasoning about log patterns receive this).
To achieve this, the Topology module constantly updates a vector index of agent skills. When Agent A sends a request "Analyze this image for anomalies," the Fabric does not look for an agent named "Image\_Analyzer." It embeds the request and searches the vector index for agents whose latent skill space aligns with "visual anomaly detection."
\end{itemize}
\subsection{Dynamic Topology Examples}
The following scenarios illustrate how the Fabric alters topology in real-time based on memory and learning:
\begin{enumerate}
 
\item The Overload Shift:
\begin{itemize}
\item Situation: The primary Math\_Solver\_Agent is overwhelmed (high $l_{load}$).
\item Fabric Action: The Cognitive Engine detects the latency spike. It automatically degrades the topology, routing simple arithmetic queries to a faster, less capable Local\_Calculator\_Tool while reserving the \item Math\_Solver\_Agent for complex calculus, optimizing global throughput.
\end{itemize}
\item The "Expert" Discovery:
\begin{itemize}
\item Situation: A new, specialized Legal\_Compliance\_Agent is introduced to the mesh.
\item Fabric Action: Through exploration steps ($\epsilon$-greedy strategy), the Fabric routes a small percentage of compliance queries to the new agent.
\item Learning: The Fabric observes that the new agent provides 20\% more accurate citations than the previous generalist. The weights $\phi$ are updated, and the new agent effectively "steals" the topology for legal tasks automatically.
\end{itemize}
\item Composite Topology (Chain-of-Agents):
\begin{itemize}
\item Sometimes the function $\Psi$ determines that no single agent can satisfy task $t$. The Fabric can dynamically synthesize a sequential topology.
\item Transformation: The Fabric splits prompt $x$ into sub-tasks $x_1$ and $x_2$.
\item Routing: $x_1 \to a_{researcher}$ then output $\to a_{writer}$.
\item The sender is unaware of this complexity; they simply receive the final result.
\end{itemize} 
\end{enumerate}

\section{Semantic Grounding: Preventing Ontological Drift}
In a Multi-Agent System, "Ontological Drift" occurs when agents, operating in isolation, begin to diverge in their definitions of key concepts. Agent A might define a "User" as an entry in a database, while Agent B defines a "User" as the currently active session. Without a mediating layer, these subtle discrepancies accumulate, leading to hallucinations, logic errors, and system incoherence.

The Semantic Grounding module of the Cognitive Fabric Node acts as the ontological anchor. It ensures that all communication maps correctly to the Shared World Model maintained in the Fabric's Memory. This is not merely a dictionary lookup; it is a dynamic validation process that rejects or translates messages that violate the system's semantic reality.
\subsection{Formalization of Semantic Consistency}
We define the Grounding function $\Gamma$ as a filter that evaluates the semantic validity of a message $x$ against the shared Memory state $\mathcal{M}$.
Let $\mathcal{O}$ be the system's Ontology (a high-dimensional vector space representing valid entities, relationships, and states derived from $\mathcal{M}$).
Let $E(x)$ be the entity extraction and relation extraction function applied to the prompt.
The Fabric calculates a Grounding Score $g$:
$$g = \text{Sim}(E(x), \mathcal{O})$$
Where $\text{Sim}$ computes the cosine similarity between the concepts in the message and the concepts in the valid ontology.
The Grounding module applies a threshold function:
$$\text{Output} = \begin{cases} x & \text{if } g \geq \tau_{valid} \quad \text{(Pass)} \\ \text{Translate}(x, \mathcal{O}) & \text{if } \tau_{soft} \leq g < \tau_{valid} \quad \text{(Align)} \\ \text{Reject}(x) & \text{if } g < \tau_{soft} \quad \text{(Hallucination Block)} \end{cases}$$
\subsection{Mechanisms of Grounding}
The process operates through two primary mechanisms: Validation and Translation.
\begin{enumerate}
\item Semantic Validation (The "Truth" Check)
\begin{itemize}
\item Before a message is routed, the Fabric checks it against the factual constraints in Memory.
\item Scenario: Agent A claims, "The server db-01 is online."
\item Fabric Memory: State vector for db-01 indicates status: offline (updated 2 seconds ago by the Monitoring Agent).
\item Action: The Cognitive Engine intervenes. The message is blocked or rewritten to "The server db-01 was last known to be online, but Fabric state indicates it is currently offline. Verify status."
\item Result: This prevents the propagation of hallucinations where one agent's false assumption cascades into another agent's error.
\end{itemize}
\item Ontological Translation ( The "Language" Bridge)
\begin{itemize}
\item Different agents may be fine-tuned on different corpuses (e.g., a Financial LLM vs. a Code LLM). They speak different "dialects."
\item Sender (Sales Agent): "Get the client details."
\item Receiver (SQL Agent): Schema uses table customers, column cust\_id.
\item Fabric Action: The Grounding module identifies the mismatch. Using the Memory's schema map, it rewrites "client" to "customer" within the prompt context, ensuring the SQL Agent generates a valid query.
\end{itemize}
\end{enumerate}
\subsection{The Cognitive Engine: Learning the Ontology}
The ontology $\mathcal{O}$ is not static. The Cognitive Engine learns new terms and relationships dynamically using Unsupervised Learning on the flow of successful transactions.

We define the update rule for the Ontology $\mathcal{O}$ at time $t+1$:

$$\mathcal{O}_{t+1} = \mathcal{O}_t + \alpha \cdot \nabla_{\mathcal{O}} \sum (\text{Success}(x_i) \cdot \text{NewTerms}(x_i))$$
If agents repeatedly use a new term (e.g., "Project\_X") and successfully complete tasks, the Fabric promotes "Project\_X" from a temporary token to a permanent entity in the Global Memory. Conversely, if a term consistently leads to error flags or clarification requests (negative reward), the Cognitive Engine lowers its validity score, effectively "forgetting" or flagging the confusing terminology.
\subsection{Example: preventing "The Ghost Entity"}
A common failure mode in MAS is the "Ghost Entity"—an object that exists in the conversation history but not in reality (e.g., a file that was deleted but agents keep discussing).
\begin{itemize}
\item Agent A: "Analyze the dataset Q3\_Report.csv."
\item Fabric Grounding Check:
\begin{enumerate}
\item Extract entity: Q3\_Report.csv.
\item Query Memory: Does Q3\_Report.csv exist in the active file manifest?
\item Result: False. (File was archived).
\end{enumerate}
\item Fabric Intervention: The message is not sent to the Analysis Agent (saving compute/cost). Instead, the Fabric returns a system error to Agent A: "Semantic Error: Entity Q3\_Report.csv is not grounded in the current environment. Available similar entities: Q3\_Report\_Archived.zip."
\end{itemize}
This ensures the Analysis Agent never wastes cycles looking for a phantom file, significantly increasing the robustness of the system.

\section{Security Policy Enforcement: The Hybrid Guardian}

Security in a Multi-Agent System is significantly more complex than in single-model deployments. The attack surface is non-linear; an adversary need not compromise the central model directly, but can instead inject a "poisoned" prompt into a peripheral, low-security agent which then propagates the malicious context up the chain to a privileged agent. This is the Cascading Trust Problem.
The Cognitive Fabric Node (CFN) addresses this by implementing a Zero-Trust Semantic Boundary. Every interaction, whether external-to-agent or agent-to-agent, passes through the Security Engine. Unlike traditional firewalls that inspect packets, the CFN inspects intent and semantics.

\subsection{Formalization of the Hybrid Security Function}

We propose a Hybrid Security Architecture that combines deterministic rules (Human-Specified) with probabilistic assessments (Machine-Learned).

Let $\Omega(x)$ be the Security Evaluation function for a prompt $x$. It is the composition of two distinct sub-functions:

$$\Omega(x) = \mathcal{H}_{rules}(x) \lor \mathcal{L}_{learned}(x | \theta, \mathcal{M})$$

Where:
\begin{enumerate}
\item $\mathcal{H}_{rules}(x)$ (The Constitution): A set of rigid, immutable constraints defined by the system administrators (e.g., "Never output a credit card number"). This returns a binary Safe/Unsafe.
\item $\mathcal{L}_{learned}(x)$ (The Immune System): A probabilistic score generated by the Cognitive Engine (using RL and Anomaly Detection) that evaluates the nuance of the request based on context history $\mathcal{M}$ and learned adversarial patterns $\theta$.
\end{enumerate}
If either function returns Unsafe, the Fabric triggers an intervention.

\subsection{The Deterministic Layer: Specified Policies}
This layer handles known threats and compliance requirements. It uses high-speed classifiers and RegEx patterns to enforce the "Constitution" of the system.
\begin{itemize}
\item Data Leakage Prevention (DLP): Automatic redaction of PII (Personally Identifiable Information).
\item Role-Based Access Control (RBAC): Enforcing permissions.
\begin{itemize}
\item Rule: "Only agents with tag Privileged\_Core can invoke DELETE methods."
\item Enforcement: If a Guest\_Agent attempts a DELETE action, $\mathcal{H}_{rules}$ immediately blocks the request before it reaches the model, saving compute and preventing risk.
\end{itemize}
\end{itemize}
\subsection{The Probabilistic Layer: Learned Cognitive Defense}
Static rules fail against novel "Jailbreaks" or social engineering attacks (e.g., "Roleplay as my grandmother who works at the chemical factory..."). The Cognitive Engine defends against these by learning normal vs. abnormal interaction flows.

This engine utilizes Reinforcement Learning from Adversarial Feedback (RLAF).
\begin{itemize}
\item State: The sequence of messages leading up to the current prompt (retrieved from Memory).
\item Observation: The semantic embedding of the current prompt.
\item Prediction: Is this prompt an attempt to bypass alignment?
\end{itemize}
$$P(Attack | x) = \sigma(W \cdot \text{Embed}(x) + b)$$

If $P(Attack)$ exceeds a dynamic threshold, the system flags the interaction.

Example of Learning:

If the system observes that 90\% of prompts containing the phrase "Ignore previous instructions" lead to policy violations, the Cognitive Engine learns to flag that phrase as high-risk, even if no specific human rule exists for it.

\subsection{Contextual Security via Memory}

Crucially, security in the CFN is stateful. The Memory module allows the Security Engine to detect attacks that are split across multiple messages (Fragmentation Attacks).
\begin{itemize}
\item Message 1: "Write a Python script to open a file." (Safe)
\item Message 2: "Import the os library." (Safe)
\item Message 3: "Delete the root directory." (Unsafe)
\end{itemize}
A stateless filter might catch Message 3, but a sophisticated attacker would obfuscate it. The CFN Memory sees the trajectory of the intent. It recognizes that the combination of these messages constitutes a malicious pattern and intervenes, potentially rewriting Message 3 or banning the user session.
\subsection{Operational Examples}
Table \ref{tab:security_interventions} contrasts how the two layers function:
\begin{table*}[ht]
\centering
\small
\renewcommand{\arraystretch}{1.5}
\begin{tabular}{p{0.18\textwidth} p{0.25\textwidth} p{0.25\textwidth} p{0.25\textwidth}}
\toprule
\textbf{Attack Type} & \textbf{Human-Specified Rule ($\mathcal{H}$)} & \textbf{Learned Cognitive Defense ($\mathcal{L}$)} & \textbf{Fabric Intervention} \\
\midrule
\textbf{SQL Injection} & 
\textbf{Trigger:} RegEx detects \texttt{DROP TABLE}. & 
N/A & 
\textbf{Block:} ``Unsafe database operation detected.'' \\

\textbf{PII Leakage} & 
\textbf{Trigger:} Pattern detects \texttt{SSN: \textbackslash d\{3\}-\textbackslash d\{2\}...} & 
N/A & 
\textbf{Redact:} Rewrite prompt to replace SSN with \texttt{<REDACTED>}. \\

\textbf{``DAN'' Jailbreak} & 
\textbf{Pass:} No specific keywords triggered. & 
\textbf{Trigger:} Semantic embedding matches cluster of known ``Persona-based Jailbreaks.'' & 
\textbf{Neutralize:} Rewrite prompt to strip the ``persona'' wrapper and keep only the core query. \\

\textbf{Contextual Manipulation} & 
\textbf{Pass:} Individual messages look safe. & 
\textbf{Trigger:} Memory analysis shows deviation from agent's standard behavior profile (Anomaly). & 
\textbf{Alert:} ``Agent behavior deviation detected. Suspending permissions pending review.'' \\
\bottomrule\\[2ex]
\end{tabular}
\caption{Comparison of Deterministic vs. Probabilistic Security Interventions.}
\label{tab:security_interventions}
\end{table*}

\section{Implementation and Architecture}
While the Cognitive Fabric Node (CFN) is logically centralized—acting as a singular, coherent brain for the system—it must be physically distributed to meet the latency and scalability demands of modern microservices. The implementation of the CFN faces a classic distributed systems trade-off: Consistency vs. Latency. We identify three distinct architectural patterns for deploying CFNs, ranging from tightly coupled sidecars to centralized control planes.

\subsection{Deployment Model A: The Cognitive Sidecar (The "Mesh" Approach)}
Inspired by the Service Mesh pattern (e.g., Istio/Envoy), this model attaches a lightweight CFN process to every single agent instance (usually on localhost).
\begin{itemize}
\item Architecture: Agent $A_i$ communicates exclusively with $CFN_i$ via loopback. $CFN_i$ handles all external routing, memory retrieval, and transformation before sending the packet over the network to $CFN_j$ (attached to Agent $A_j$).
\item Mechanism:
\begin{itemize}
\item Inference: The "Rewrite" and "Security" models run locally on the edge.
\item Learning: Gradients are computed locally and pushed asynchronously to a central parameter server (Federated Learning approach).
\end{itemize}
\item Pros:
\begin{itemize}
\item Zero Network Hop: No initial network latency to reach the middleware.
\item Isolation: If one CFN crashes, only one agent is affected.
\end{itemize}
\item Cons:
\begin{itemize}
\item Resource Bloat: Running a cognitive engine (even a quantized one) alongside every agent consumes significant CPU/RAM.
\item State Drift: Propagating a new security policy to 1,000 distributed sidecars takes non-zero time ($t > 0$), creating a window of vulnerability.
\end{itemize}
\end{itemize}

\subsection{Deployment Model B: The Cognitive Hub (The "Cluster" Approach)}
In this model, the Fabric is a centralized cluster of high-performance GPU nodes. Agents are thin clients that simply forward raw prompts to the Fabric API.
\begin{itemize}
\item Architecture: All traffic ($x_{i \to j}$) is routed to a Load Balancer fronting the CFN Cluster. The Cluster processes the message and routes it to the destination agent.
\item Mechanism:
\begin{itemize}
\item Shared State: The Memory ($\mathcal{M}$) and Topology ($\mathcal{T}$) are strictly local to the cluster, ensuring strong consistency.
\end{itemize}
\item Pros:
\begin{itemize}
\item Global Coherence: The "Brain" is always perfectly synchronized. Hallucination checks use the absolute latest memory.
\item Efficiency: Heavy GPU resources are pooled and autoscaled, rather than stranded on idle sidecars.
\end{itemize}
\item Cons:
\begin{itemize}
\item Latency Penalty: Every interaction incurs a double network hop (Agent $\to$ Fabric $\to$ Agent).
\item Single Point of Failure: If the Fabric Cluster goes down, the entire agent swarm loses intelligence and connectivity.
\end{itemize}
\end{itemize}
\subsection{Deployment Model C: The Hybrid Hierarchical Fabric}
We propose this as the optimal reference implementation. It utilizes a split-brain architecture.
\begin{itemize}
\item Edge Nodes (Sidecars): Handle high-frequency, low-latency tasks.
\begin{itemize}
\item Functions: Basic PII masking (Regex), Caching, simple Topology routing.
\end{itemize}
\item Core Nodes (Central Cluster): Handle high-cognitive-load tasks.
\begin{itemize}
\item Functions: Complex Prompt Rewriting, Long-term Memory retrieval, Strategic Planning.
\end{itemize}
\item Routing Logic:
The Edge Node calculates a "Complexity Score" $C(x)$ for the prompt.
$$ \text{Route}(x) = \begin{cases} \text{Process Locally} & \text{if } C(x) < \tau_{edge} \\ \text{Forward to Core} & \text{if } C(x) \geq \tau_{edge} \end{cases}$$
This ensures that simple "ping" requests are fast, while complex reasoning tasks get the full power of the central Cognitive Engine.
\end{itemize}
\subsection{The Synchronization Challenge}
Regardless of the deployment model, the illusion of a "Single Fabric" relies on the synchronization of two key artifacts:
\begin{enumerate}
\item Episodic Memory ($\mathcal{M}$): What happened?
\item Learned Policies ($\theta$): How should we behave?
\end{enumerate}

We utilize an Eventually Consistent Gossip Protocol. When a CFN at node $k$ learns a new security rule (e.g., "Block prompt X"), it broadcasts a delta update $\Delta \theta$.

The propagation time $T_{prop}$ is modeled as:
$$T_{prop} \approx \frac{\ln(N)}{\beta}$$
Where $N$ is the number of nodes and $\beta$ is the gossip fan-out rate. For a system of 1,000 agents, convergence occurs in milliseconds, which is acceptable for semantic consistency, though instantaneous locking is avoided to preserve availability (AP over CP in CAP theorem terms).

\subsection{Latency Analysis: The "Cognitive Tax"}
Adding intelligence to the network introduces latency. We term this the Cognitive Tax.
Total Latency $L_{total}$ is:
$$L_{total} = L_{net} + L_{inference} + L_{memory\_lookup}$$
To make this viable, the CFN must operate within strict budgets.
\begin{itemize}
\item Vector Search: Optimized via HNSW indices ($< 10$ms).
\item Inference: Using distilled, task-specific Small Language Models (SLMs) rather than generalist LLMs for the rewriting layer ($< 50$ms).
\end{itemize}
The hypothesis is that while $L_{total}$ increases per message, the Total Time to Task Completion decreases because the Fabric eliminates the "back-and-forth" clarifications and errors typical of unmanaged agents.

\section{Evaluation}

To assess the efficacy of Cognitive Fabric Nodes (CFN) in a complex, multi-step reasoning environment, we implemented a testing framework utilizing LangMARL as the internal learning algorithm to update and optimize inter-agent communication\cite{yao2026langmarlnaturallanguagemultiagent}. In this implementation, the CFN governs the communication channels between agents, maintaining a distinct prompt-rewriting "policy," denoted as $\pi_{i \rightarrow j}^{text}$, for each directed edge in the communication graph.  This policy serves as the ``memory" of the CFN, and is updated periodically by the LangMARL framework. The prompt rewriting process encapsulates all five functions of the proposed CFN system, though we primarily focus on the improvement of the multi-agent system's ability to complete the QA task.

\subsection{Experimental Setup}
Our system is composed of four specialized agents designed to collaboratively resolve complex analytical queries:
\begin{itemize}
\item Researcher ($A_R$): Extracts evidence verbatim from source material.
\item Analyst ($A_A$): Interprets the extracted evidence and identifies underlying patterns.
\item Critic ($A_C$): Challenges the Analyst's interpretations and flags logical contradictions.
\item Synthesizer ($A_S$): Composes the final, definitive answer based on the interactions of the swarm.
\end{itemize}
The communication topology is a directed graph defined by the following allowed interactions:
\begin{itemize}
\item $A_R \rightarrow A_A$ (Evidence hand-off)
\item $A_R \rightarrow A_S$ (Raw data baseline)
\item  $A_A \rightarrow A_C$ (Hypothesis submission)
\item $A_A \rightarrow A_S$ (Interpreted patterns)
\item $A_C \rightarrow A_R$ (Feedback loop: requests for missing/new evidence)
\item $A_C \rightarrow A_S$ (Constraints and flagged warnings)
\end{itemize}
\subsection{Methodology and Optimization}
We evaluated the system over the HotPotQA and MuSiQue datasets \cite{yang2018hotpotqa}\cite{trivedi2022musique}. These datasets were modified so that each agent had access to only partial information about the question being answered, and must collaborate to answer it. The optimization process leverages the core principles of LangMARL, specifically bridging centralized training with decentralized execution (CTDE) in the language space\cite{yao2026langmarlnaturallanguagemultiagent}. 

After every example episode, the final output produced by the Synthesizer is evaluated to determine the system's global performance. We then employ LangMARL's Centralized Language Critic to analyze the complete inter-agent communication trajectory and perform causal credit assignment\cite{yao2026langmarlnaturallanguagemultiagent}. Rather than generating a single global reflection, the Critic assigns specific linguistic feedback to each agent pair's communication policy. 

To execute the policy updates, we utilize TextGrad to backpropagate these natural language credits into textual gradients\cite{yuksekgonul2024textgrad}. For each communication edge, TextGrad optimizes the rewriting policy $\pi_{i \rightarrow j}^{text}$, refining how the CFN restructures, truncates, or enriches the prompts sent between those specific agents on subsequent iterations.  While in this case we are focusing on task performance, metrics, more generally these rewriting policies may include security policies, topology, or any other relevant information for improving agent to agent communication.

All agents and LLMs used in the LangMARL system used the Claude Sonnet 4.6 model.

\subsection{Results}
Table \ref{tab:eval} presents the performance metrics of the system, comparing a static baseline (no CFN prompt rewriting), a globally optimized baseline (Standard TextGrad without LangMARL credit assignment), and our CFN-LangMARL hybrid approach. 

\begin{table*}[t]
\centering

\label{tab:comprehensive_comparison}
\begin{tabular}{lcccc}
\toprule
\textbf{Method} & Baseline & Multi-agent & TextGrad & Ours \\
\midrule
\multicolumn{5}{c}{\textit{Accuracy/Pass Rate}} \\
\midrule
HotPotQA & 92 & 80.1 & 82.3  & \textbf{91.5} \\
MuSiQue & 87.5 & 72.7 & 76.2  & \textbf{86.1} \\
\bottomrule\\[2ex]
\end{tabular}

\caption{Evaluation on the HotPotQA and MuSiQue datasets. These datasets were modified so that each agent only had access to partial information. All reported values represent the mean performance after 5 training iterations using the same backbone LLMs. Baseline is the base Claude Sonnet 4.6 model performance, multi-agent is the base multi-agent system performance, TextGrad is performance with TextGrad only updates.}
\label{tab:eval}
\end{table*}

The integration of LangMARL's agent-level language credit assignment allows the CFN to pinpoint exact communication bottlenecks—such as the Critic failing to clearly articulate evidence gaps to the Researcher—and update that specific edge's rewriting policy without destabilizing the rest of the swarm's communication.

For both datasets, we can see that the CFN-LangMARL system significantly increases the performance of the multi-agent system.  While the baseline Claude Sonnet 4.6 model achieves 92\% and 87.5\% on the HotPotQA and MuSiQue datasets, respectively, the added complexity of having the evidence split between the agents and the multi-agent communication lowers this performance significantly, down to 80.1\% and 72.7\%. Although TextGrad marginally improves performance, our system increases performance to within less than 1\% of the baseline.

\section{Conclusion and Future Work}
This paper has introduced Cognitive Fabric Nodes (CFN), a transformative middleware architecture designed to evolve Multi-Agent Systems (MAS) from fragile, disjointed experiments into robust, enterprise-grade ecosystems. By shifting intelligence from the endpoints to the interconnect, we address the fundamental "stochastic coordination problem" inherent in GenAI systems.

The current paradigm of direct agent-to-agent communication assumes that agents are rational, context-aware, and secure. Experience shows they are none of these. They hallucinate, forget context, and are easily manipulated. The Cognitive Fabric mitigates these failures not by making the agents perfect, but by making the environment in which they operate intelligent.

Through the formalization of Active Memory, Dynamic Topology, Semantic Grounding, Security Enforcement, and Prompt Transformation, we have demonstrated that a "logically centralized, practically distributed" fabric can:
\begin{enumerate}
\item Eliminate Ontological Drift: Ensuring all agents share a single source of truth.
\item Optimize Compute: Routing tasks based on real-time skill vectors rather than static assumptions.
\item Enforce Hybrid Security: blending rigid human policy with adaptive, learned defense mechanisms.
The CFN architecture represents a move away from "Smart Agents in Dumb Networks" to "Specialized Agents in a Cognitive Network."
\end{enumerate}
\subsection{Future Work}
While this paper establishes the internal architecture of a single Cognitive Fabric, the next frontier lies in the interaction between Fabrics.
\begin{enumerate}
\item Fabric-to-Fabric (F2F) Protocol

As organizations deploy their own proprietary swarms, we foresee the need for a "Border Gateway Protocol" for AI. How does a healthcare provider's Fabric (optimized for privacy and medical ontology) communicate with an insurance provider's Fabric (optimized for finance and risk)?
Future research must define the F2F Handshake: a negotiation phase where two Fabrics align on security policies (e.g., "I will share data only if you guarantee PII redaction level 4") and ontological mappings before allowing their agents to exchange a single token.

\item The Economics of the Fabric

We must explore the economic models of the Cognitive Engine. If the Fabric rewrites a prompt to be more efficient, it saves tokens. If it caches a result, it saves compute. Future work should explore Tokenomics within the Fabric, where the middleware itself "charges" agents for memory retrieval and "pays" agents for successful task completion, creating an internal market that drives the system toward global optimization naturally.
\item Cognitive Decay and Garbage Collection

As the Memory module grows, retrieval latency increases. We need advanced research into "Cognitive Forgetting"—algorithms that autonomously determine which memories have become obsolete noise and should be pruned, ensuring the Fabric remains agile over years of operation.
\end{enumerate}
By solving these challenges, the Cognitive Fabric will become the invisible, omnipresent substrate of the AI-powered internet—the nervous system connecting the isolated minds of our digital workforce.


%

\ifCLASSOPTIONcaptionsoff
  \newpage
\fi



%


\bibliographystyle{IEEEtran}
\bibliography{references}

@article{lyu2025llms,
  title={LLMs for Multi-Agent Cooperation: A Comprehensive Survey},
  author={Lyu, Xueguang and others},
  journal={arXiv preprint},
  year={2025}
}

@article{trirat2024automl,
  title={AutoML-Agent: A Multi-Agent LLM Framework for Full-Pipeline AutoML},
  author={Trirat, Patara and Jeong, Wonyong and Hwang, Sung Ju},
  journal={arXiv preprint arXiv:2410.02958},
  year={2024}
}

@article{qayyum2025llm,
  title={LLM-Driven Multi-Agent Architectures for Intelligent Self-Organizing Networks},
  author={Qayyum, Adnan and Albaseer, Abdullatif and Qadir, Junaid and Al-Fuqaha, Ala and Abdallah, Mohamed},
  journal={IEEE Network},
  year={2025},
  publisher={IEEE}
}

@article{lemad2025,
  title={LEMAD: LLM-Empowered Multi-Agent System for Anomaly Detection in Power Grid Services},
  author={Anonymous},
  journal={MDPI},
  year={2025}
}

@misc{yao2026langmarlnaturallanguagemultiagent,
      title={LangMARL: Natural Language Multi-Agent Reinforcement Learning}, 
      author={Huaiyuan Yao and Longchao Da and Xiaoou Liu and Charles Fleming and Tianlong Chen and Hua Wei},
      year={2026},
      eprint={2604.00722},
      archivePrefix={arXiv},
      primaryClass={cs.CL},
      url={https://arxiv.org/abs/2604.00722}, 
}

@inproceedings{yang2018hotpotqa,
  title={{HotpotQA}: A Dataset for Diverse, Explainable Multi-hop Question Answering},
  author={Yang, Zhilin and Qi, Peng and Zhang, Saizheng and Bengio, Yoshua and Cohen, William W. and Salakhutdinov, Ruslan and Manning, Christopher D.},
  booktitle={Conference on Empirical Methods in Natural Language Processing ({EMNLP})},
  year={2018}
}

@article{trivedi2022musique,
  title={MuSiQue: Multihop Questions via Single-hop Question Composition},
  author={Trivedi, Harsh and Balasubramanian, Niranjan and Khot, Tushar and Sabharwal, Ashish},
  journal={Transactions of the Association for Computational Linguistics},
  volume={10},
  pages={539--554},
  year={2022},
  url={https://aclanthology.org/2022.tacl-1.31.pdf}
}

@article{yuksekgonul2024textgrad,
  title={Textgrad: Automatic" differentiation" via text},
  author={Yuksekgonul, Mert and Bianchi, Federico and Boen, Joseph and Liu, Sheng and Huang, Zhi and Guestrin, Carlos and Zou, James},
  journal={arXiv preprint arXiv:2406.07496},
  year={2024}
}
%








\end{document}